\documentclass[preprint2,numberedappendix]{emulateapj-rtx4}

\usepackage{bm,color}
\newcommand{\Eq}[1]{Equation~(\ref{#1})}

\begin{document}

\title{Magnetic helicity and energy spectra of a solar active region}

\author{Hongqi Zhang$^1$, Axel Brandenburg$^{2,3}$ and D.D. Sokoloff$^{4,5}$}
\affil{
$^1$Key Laboratory of Solar Activity, National Astronomical Observatories, Chinese
Academy of Sciences, Beijing 100012, China \\ \email{E-mail: hzhang@bao.ac.cn}
$^2$Nordita, KTH Royal Institute of Technology and Stockholm University,
Roslagstullsbacken 23, 10691 Stockholm, Sweden\\
$^3$Department of Astronomy, AlbaNova University Center,
Stockholm University, 10691 Stockholm, Sweden\\
$^4$Department of Physics, Moscow University, 119992 Moscow, Russia\\
$^5$Pushkov Institute of Terrestrial Magnetism, Ionosphere and Radiowave
Propagation of the Russian Academy of Sciences,\\
Troitsk, Moscow, 142190, Russia
}

\begin{abstract}
We compute for the first time
magnetic helicity and energy spectra of the solar
active region NOAA 11158 during 11--15 February 2011
at $20^\circ$ southern heliographic latitude
using observational photospheric vector magnetograms.
We adopt the isotropic representation of the Fourier-transformed
two-point correlation tensor of the magnetic field.
The sign of magnetic helicity turns out to be predominantly
positive at all wavenumbers.
This sign is consistent with what is theoretically expected
for the southern hemisphere.
The magnetic helicity normalized to its theoretical maximum value,
here referred to as {\it relative} helicity,
is around 4\% and strongest at intermediate
wavenumbers of $k\approx0.4\,{\rm Mm}^{-1}$, corresponding to a scale
of $2\pi/k\approx16\,{\rm Mm}$.
The same sign and a similar value are also found for the
relative current helicity evaluated in real space based on the
vertical components of magnetic field and current density.
The modulus of the magnetic helicity spectrum shows a $k^{-11/3}$
power law at large wavenumbers, which implies a $k^{-5/3}$ spectrum
for the modulus of the current helicity.
A $k^{-5/3}$ spectrum is also obtained for the magnetic energy.
The energy spectra evaluated separately from the horizontal
and vertical fields agree for wavenumbers below 3\,Mm$^{-1}$,
corresponding to scales above 2\,Mm.
This gives some justification to our assumption of isotropy and places
limits resulting from possible instrumental artefacts at small scales.
\end{abstract}


\section{Introduction}

Magnetic helicity is an important quantity that reflects the topology
of the magnetic field (Woltjer, 1958a,b and Taylor, 1986).
Pioneering studies of magnetic helicity in solar physics have been performed
by several authors focussing on the accumulation of magnetic helicity in the
solar atmosphere \citep[e.g.][]{Berger84,Chae01}, the
force-free $\alpha$ coefficient, and the mean current helicity density
in solar active regions \citep{Seehafer90}.

Besides the hemispheric sign distribution of large-scale helical features in
active regions \citep{Pevtsov94,Ab97},
there can be patches of right-handed and left-handed fields corresponding
respectively to positive and negative helicities, intermixed
in a mesh-like pattern in the sunspot umbra and a threaded pattern in the
sunspot penumbra \citep{Su09}.
\cite{Zhang} showed that the individual magnetic fibrils tend to be
dominated by the current density component caused by magnetic
inhomogeneity, while the large-scale magnetic region tends to be
dominated by the component of the current density associated with
the magnetic twist.
\cite{Venkatakrishnan09} pointed out that the existence of
global twist for a sunspot -- even in the absence of a net current -- is
consistent with a fibril structure of sunspot magnetic fields.

\begin{figure*}
\begin{center}
\includegraphics[width=\textwidth]{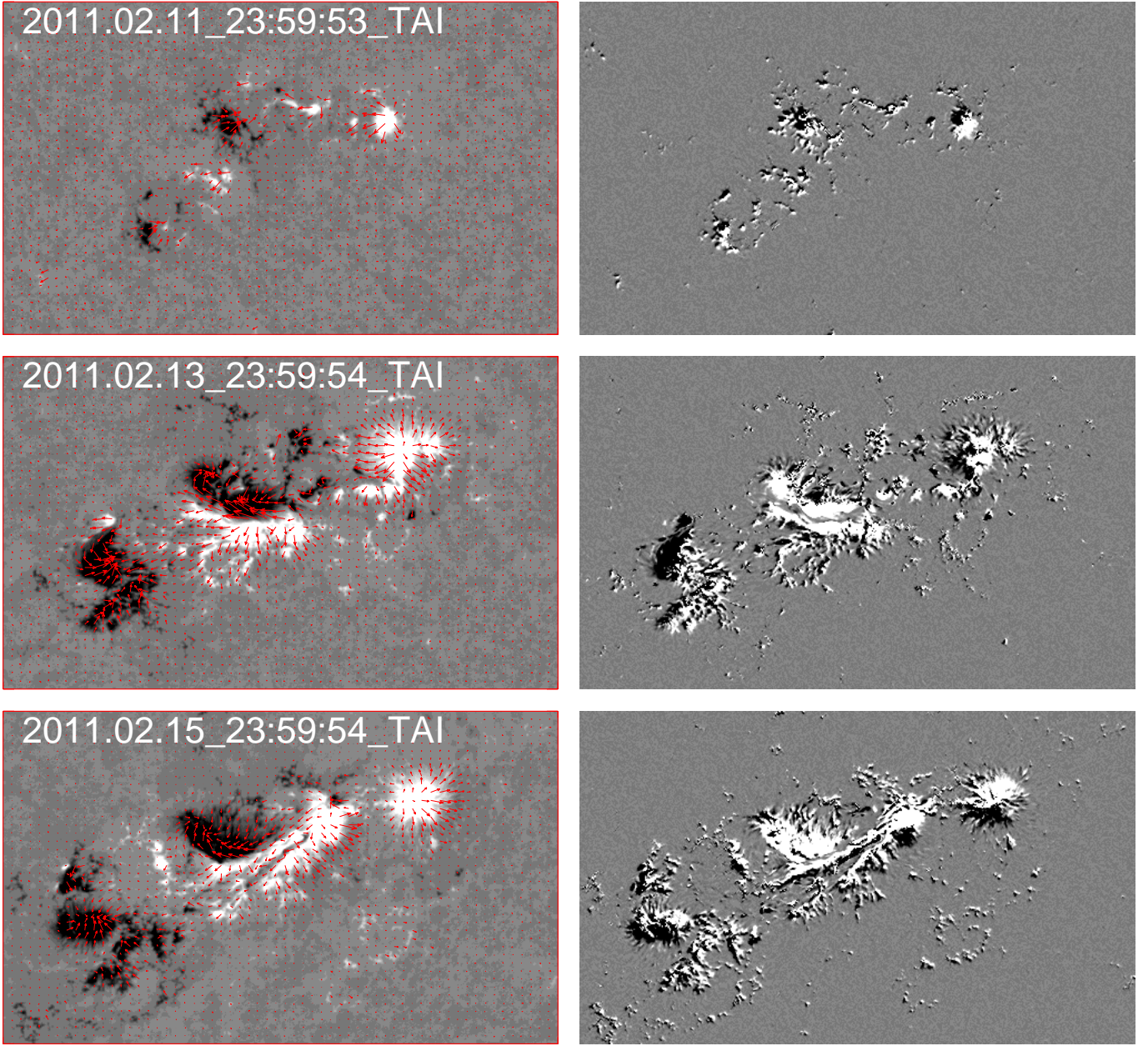}
\end{center}
\caption{
Photospheric vector magnetograms (left) and plots of $J_zB_z$ (right)
for the active region NOAA 11158 between 11--15 February 2011.
The arrows show the transverse component of the magnetic field.
Light (dark) shades indicate positive (negative) values of $B_z$
on the left and $J_zB_z$ on the right.
}\label{fig:helispec1}
\end{figure*}

The redistribution of magnetic helicity contained within different
scales was argued to be the interchange of twist and writhe due to
magnetic helicity conservation \citep[cf.][]{Zeldovich83,KB99}.
Furthermore, the spectral magnetic helicity distribution is important
for understanding the operation of the solar dynamo \citep{BS05}.
It has been argued that, if the large-scale magnetic field is generated
by an $\alpha$ effect \citep{KR80}, it must produce magnetic helicity
of opposite signs at large and small length scales \citep{See96,Ji99}.
We call such a magnetic field bi-helical \citep{YB03}.
To alleviate the possibility of catastrophic
(magnetic Reynolds number-dependent) quenching
of the $\alpha$ effect \citep{GD94} and slow saturation \citep{Bra01},
one must invoke magnetic helicity fluxes from small-scale magnetic fields
\citep{KMRS00,BF00,BS05,BCC09,HB12}.

In the present paper, we determine the spectrum of magnetic helicity
and its relationship with magnetic energy from photospheric vector
magnetograms of a solar active region.
We use a technique that is based on the spectral representation of the
magnetic two-point correlation tensor.
It is related to the method of \cite{MGS82} for determining the magnetic helicity
spectrum from in situ measurements of the magnetic field in the solar wind.
Their key assumption allowing for the determination of magnetic helicity spectra
is that of homogeneity.
This technique was recently applied to data from {\em Ulysses}
to show that the magnetic
field at high heliographic latitudes has opposite signs of helicity
in the two hemispheres and also at large and small length scales \citep{BSBG11};
see also \cite{WBM11,WBM12} for results from corresponding simulations.
In the present work, a variant is proposed where we assume
local statistical isotropy in the horizontal plane to compute
magnetic energy and helicity spectra.

\section{Data analysis}

We have analyzed data from the solar active region NOAA 11158 during
11--15 February 2011, taken by the Helioseismic and Magnetic Imager (HMI)
on board the {\em Solar Dynamics Observatory} (SDO).
The pixel resolution of the magnetogram is about $0.5''$, and the field
of view is $250''\times 150''$.
Figure \ref{fig:helispec1} shows photospheric vector magnetograms (left)
and the corresponding distribution of $h_C^{(z)}=J_z B_z$ (right)
from the vector magnetograms of that active region on different days.
Here, $J_z=\partial B_y/\partial x-\partial B_x/\partial y$, and
$J_z/\mu_0$ is the vertical component of the current density in
SI units with $\mu_0$ being the vacuum permeability, while in cgs units,
the current density is $J_z c/4\pi$ with $c$ being the speed of light.
The superscript `$(z)$' on $h_C^{(z)}$ indicates that only the vertical
contribution to the current helicity density is available.

It turns out that the mean value of the current helicity density,
${\cal H}_C^{(z)}=\langle{h_C^{(z)}}\rangle$,
is positive and $\approx2.7$\,G$^2$\,km$^{-1}$.
Furthermore, as a proxy of the force-free $\alpha$ parameter,
we determine $\alpha=J_z/B_z$, which is on the average
$\langle{\alpha}\rangle\approx2.8\times 10^{-5}$\,km$^{-1}$.
For future reference, let us estimate the current
helicity normalized to its theoretical maximum value,
henceforth referred to as {\em relative} helicity.
This is not to be confused with the gauge-invariant
magnetic helicity relative to that of an associated
potential field \citep{Berger84}.
Thus, we consider the ratio
\begin{equation}
r_C=\langle J_zB_z\rangle\left/
\left(\langle J_z^2\rangle\langle B_z^2\rangle\right)^{1/2}\right.
\end{equation}
as an estimate for the relative current helicity.
For the active region NOAA 11158 we find $r_C=+0.034$.
This value is based on one snapshot, but similar values have
been found at other times.

Let us now turn to the two-point correlation tensor,
$\langle B_i({\bm x},t)\,B_j({\bm x}+{\bm\xi},t)\rangle$, where ${\bm x}$
is the position vector on the two-dimensional surface,
and angle brackets denote ensemble averaging or, in the present case,
averaging over annuli of constant radii, i.e., $|{\bm\xi}|={\rm const}$.
Its Fourier transform with respect to ${\bm\xi}$ can be written as
\begin{equation}
\left\langle\hat{B}_i({\bm k},t)\hat{B}_j^*\!({\bm k}',t)\right\rangle
=\Gamma_{ij}({\bm k},t)\delta^2({\bm k}-{\bm k}'),
\end{equation}
where $\hat{B}_i({\bm k},t)=\int B_i({\bm x},t)
\,e^{i{\bm k}\cdot{\bm x}}d^2x$ is the two-dimensional Fourier transform,
the subscript $i$ refers to one of the three magnetic field components,
the asterisk denotes complex conjugation,
and ensemble averaging will be replaced by averaging over
concentric annuli in wavevector space.
Following \cite{MGS82}, it is possible to determine the magnetic helicity
spectrum from the spectral correlation tensor $\Gamma_{ij}({\bm k},t)$
by making the assumption of local statistical isotropy.
At the end of this paper we consider the applicability of this
assumption in more detail.
Considering that $\bm{k}$ defines the only preferred direction in $\Gamma_{ij}$,
and that $k_i\hat{B}_i=0$, the only possible structure of
$\Gamma_{ij}({\bm k},t)$ is \citep[cf.][]{Moffatt78}
\begin{equation}
\Gamma_{ij}(\bm{k},t)=\frac{2E_M(k,t)}{4\pi k}(\delta_{ij}-\hat{k}_i\hat{k}_j)
+\frac{iH_M(k,t)}{4\pi k}\varepsilon_{ijk}k_k,\label{eq:helispec5}
\end{equation}
where $\hat{k}_i=k_i/k$ is a component of the unit vector of $\bm{k}$,
$k=|\bm{k}|$ is its modulus with $k^2=k_x^2+k_y^2$,
and $E_M(k,t)$ and $H_M(k,t)$ are the magnetic energy and
magnetic helicity spectra\footnote{We use this opportunity to point out
a sign error in the corresponding Equation~(3) of \cite{BSBG11}.
Their results were however based on the equation $H_M(k)=4\mbox{Im}
\langle\hat{B}_T\hat{B}_N^*\!\rangle$, which has the correct sign.
Here, $\hat{B}_T$ and $\hat{B}_N$ are transverse and normal components
of the Fourier-transformed magnetic field.}, normalized such that
\begin{eqnarray}
{\cal E}_M(t)\equiv
\frac{1}{2}\langle{\bm{B}^{2}}\rangle\!&=&\!{\int}^\infty_0\!\! E_M(k,t)\,dk\nonumber ,\\
{\cal H}_M(t)\equiv
\langle{\bm{A}\cdot\bm{B}}\rangle\!&=&\!\int^\infty_0\!\!
H_M(k,t)\,dk.\label{eq:helispec6}
\end{eqnarray}
Note that the mean energy density in ${\rm erg}/{\rm cm}^{3}$
is ${\cal E}_M/4\pi$.
We emphasize that the expression for $\Gamma_{ij}({\bm k},t)$ differs
from that of \cite{Moffatt78} by a factor $2k$, because we are here in
two dimensions, so the differential for the integration over shells in
wavenumber space changes from $4\pi k^2\,dk$ to $2\pi k\,dk$.

Note that the magnetic vector potential is not an observable quantity,
so the magnetic helicity might not be gauge-invariant.
However, if the spatial average is over all space, or if the
magnetic field falls off sufficiently rapidly toward the boundaries,
both ${\cal H}_M(t)$ and $H_M(k,t)$ are gauge-invariant.
Indeed, with the present analysis, $H_M(k,t)$ is manifestly gauge-invariant,
because it has been computed directly from the magnetic field as obtained
through the photospheric vector magnetogram.

The components of the correlation tensor of the turbulent magnetic
field can be written in the form
\begin{eqnarray}
\label{eq:helitens}
&&4\pi k{\bf \Gamma}(k,\phi_k)=\\
&&\left(\begin{array}{ccc} (1-\cos^2\phi_k)2E_M & -\sin2\phi_kE_M & -ik\sin\phi_k H_M\\
-\sin2\phi_kE_M & (1-\sin^2\phi_k)2E_M & \;\;ik\cos\phi_k H_M\\
ik\sin\phi_k H_M &  -ik\cos\phi_k H_M & 2E_M 
\end{array}\right),\nonumber
\end{eqnarray}
where we have defined the polar angle in wavenumber space,
$\phi_k = {\rm Arctan}(k_y, k_x)$,
so that $k_x=k\cos\phi_k$ and $k_y=k\sin\phi_k$.
For brevity, we have also skipped the arguments $k$ and $t$ on
$E_M(k,t)$ and $H_M(k,t)$.

In the following we present shell-integrated spectra.
However, because we consider here two-dimensional spectra,
they correspond to the power in annuli of radius $k$ and are obtained as
\begin{eqnarray}
2E_M(k,t)&=&2\pi k\,\mbox{Re}\left\langle \Gamma_{xx}+\Gamma_{yy}+\Gamma_{zz}\right\rangle_{\phi_k},\\
kH_M(k,t)&=&4\pi k\,\mbox{Im}\left\langle\cos\phi_k\Gamma_{yz}-\sin\phi_k\Gamma_{xz}\right\rangle_{\phi_k},
\end{eqnarray}
where the angle brackets with subscript $\phi_k$
denote averaging over annuli in wavenumber space.

\begin{figure}
\begin{center}
\includegraphics[width=\columnwidth]{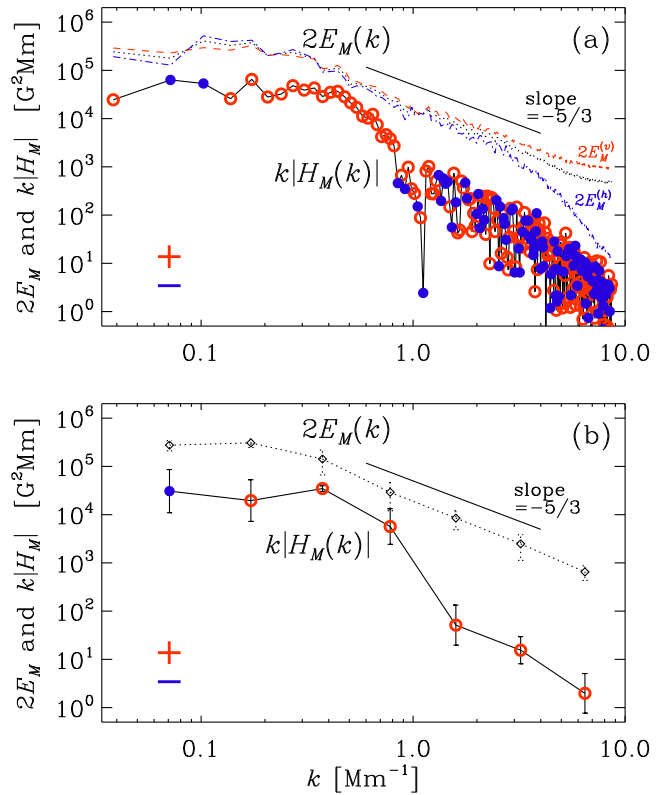}
\end{center}
\caption{(a) $2E_M(k)$ (dotted line) and $k|H_M(k)|$ (solid line)
for NOAA 11158 at 23:59:54\,UT on 13 February 2011.
Positive (negative) values of $H_M(k)$ are indicated by open (closed)
symbols, respectively.
$2E_M^{(h)}(k)$ (red, dashed) and $2E_M^{(v)}(k)$ (blue, dash-dotted) are
shown for comparison.
(b) Same as upper panel, but the magnetic helicity is averaged
over broad logarithmically spaced wavenumber bins.
\label{fig:phelicity}
}\end{figure}

The realizability condition \citep{Moffatt69} implies that
\begin{equation}
k|H_M(k,t)|\le2E_M(k,t).
\label{realizability}
\end{equation}
It is therefore convenient to plot $k|H_M(k,t)|$ and $2E_M(k,t)$
on the same graph, which allows one to judge how helical
the magnetic field is at each wavenumber.
Furthermore, to assess the degree of isotropy, we also consider magnetic
energy spectra $E_M^{(h)}(k)$ and $E_M^{(v)}(k)$ based respectively on the horizontal
and vertical magnetic field components, defined via
\begin{eqnarray}
2E_M^{(h)}(k)&=&4\pi k\,\mbox{Re}\left\langle \Gamma_{xx}+\Gamma_{yy}\right\rangle_{\phi_k},\\
2E_M^{(v)}(k)&=&4\pi k\,\mbox{Re}\left\langle \Gamma_{zz}\right\rangle_{\phi_k}.
\end{eqnarray}
Under isotropic conditions, we expect $E_M(k)\approx E_M^{(h)}(k)\approx E_M^{(v)}(k)$.

We now consider magnetic energy and helicity spectra for the
active region NOAA 11158.
The calculated region of the field of view is $256''\times 256''$,
i.e.\ $512\times 512$ pixels or $L^2=(186\,{\rm Mm})^2$.
We present first the results for NOAA 11158 at 23:59:54UT on 13 February 2011;
see Figure~\ref{fig:phelicity}(a).
It turns out that the magnetic energy spectrum has a clear $k^{-5/3}$ range
for wavenumbers in the interval $0.5\,{\rm Mm}^{-1}<k<5\,{\rm Mm}^{-1}$.
The magnetic helicity spectrum is predominantly positive at intermediate
wavenumbers, but we also see that toward high wavenumbers the magnetic
helicity is fluctuating strongly around small values.
To determine the sign of magnetic helicity at these smaller scales,
we average the spectrum over broad, logarithmically spaced wavenumber bins;
see the lower panel of Figure~\ref{fig:phelicity}.
This shows that even at smaller length scales the magnetic helicity is
still positive, again consistent with the fact that this active region
is at southern latitudes.

\begin{figure}
\begin{center}
\includegraphics[width=\columnwidth]{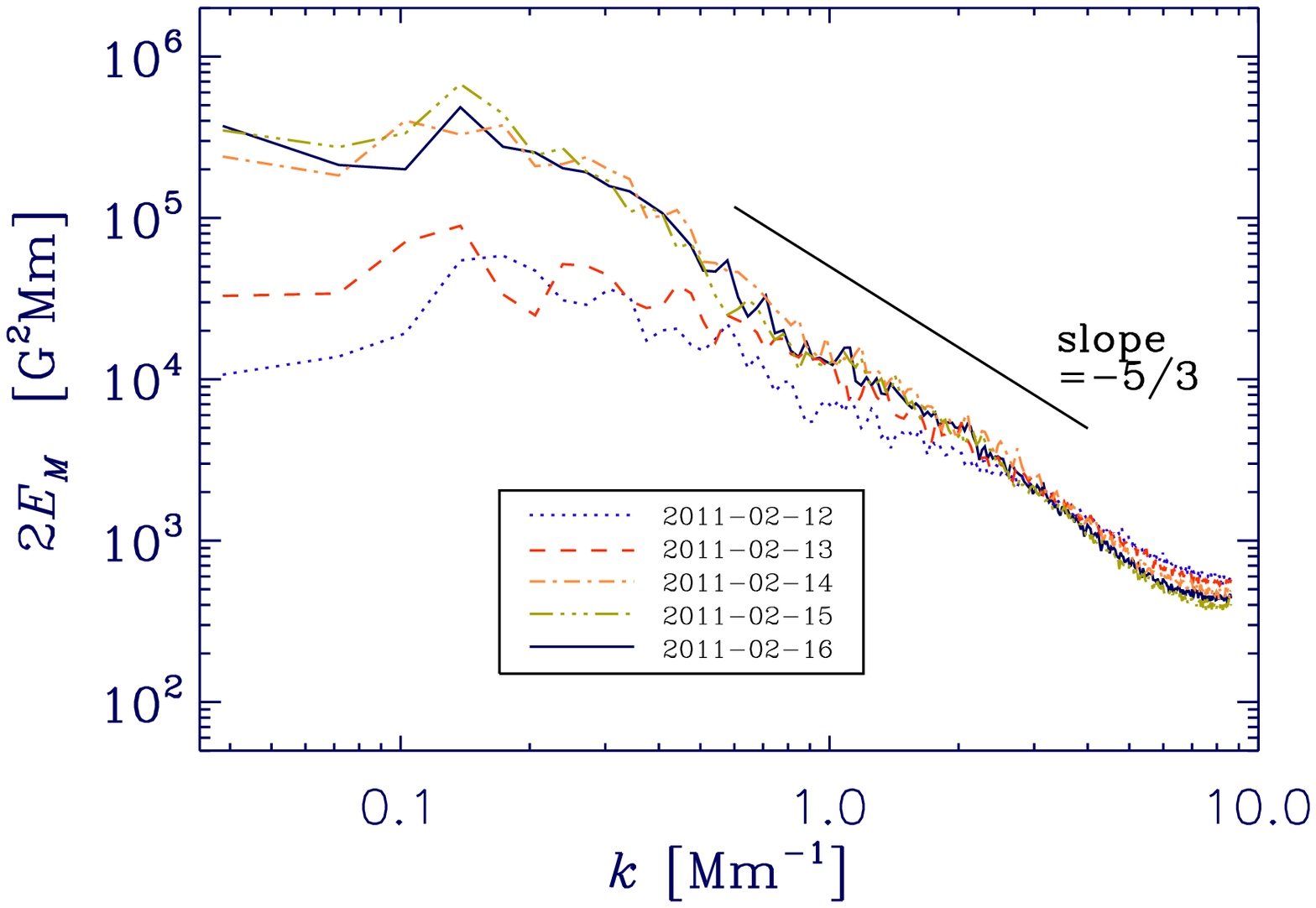}
\includegraphics[width=\columnwidth]{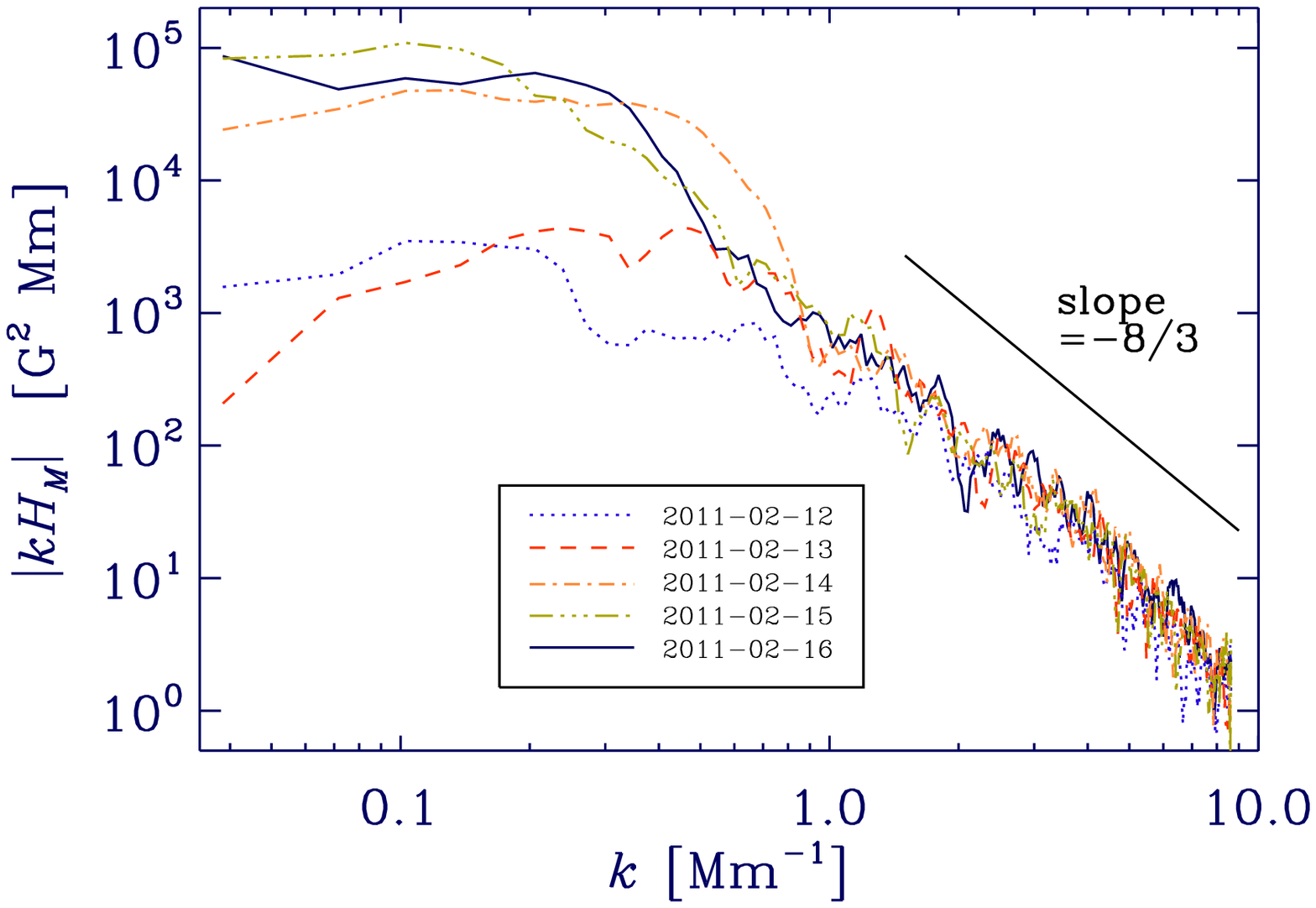}
\end{center}
\caption{
Similar to Figure~\ref{fig:phelicity}, showing $E_M(k,t)$ (upper panel)
and $k|H_M(k,t)|$ (lower panel) for the other days.
\label{fig:heli_phelicity_Bran_zhq5_ene}
}\end{figure}

To calculate the relative magnetic helicity $r_M$,
we define the integral scale of the magnetic field in the usual way as
\begin{eqnarray}
\ell_M=\left.\int k^{-1}E_M(k)\,dk\right/\int E_M(k)\,dk.
\end{eqnarray}
The realizability condition of \Eq{realizability} can be rewritten
in integrated form \citep[e.g.][]{KTBN13} as
\begin{eqnarray}
|{\cal H}_M|=\left|\int\! H_M dk\right|
\le 2\!\int \!k^{-1}E_M(k) dk \equiv 2\ell_M{\cal E}_M.\quad
\end{eqnarray}
In particular, we have $|{\cal H}_M(t)|\le 2\ell_M{\cal E}_M(t)$.
This gives
\begin{eqnarray}
r_M={\cal H}_M/2\ell_M{\cal E}_M,
\end{eqnarray}
which obeys $|r_M|\le1$.
Again, this quantity is not to be confused with the
gauge-invariant helicity of \cite{Berger84}.
For the active region NOAA 11158 at 23:59:54\,UT on 13 February 2011
we have $\ell_M\approx5.8\,{\rm Mm}$,
${\cal H}_M\approx3.3\times10^{4}\,{\rm G}^2\,{\rm Mm}$,
and ${\cal E}_M\approx6.7\times10^{4}\,{\rm G}^2$, so $r_M\approx0.042$.
The relative magnetic helicity has thus the same sign as the relative
current helicity.
The corresponding magnetic column energy in the two-dimensional
domain of size $L^2$ is
$L^2{\cal E}_M/4\pi\approx1.8\times10^{24}\,{\rm erg}\,{\rm cm}^{-1}$,
which is about three times larger than the values given by \cite{SZYL13}.
The magnetic column helicity is
$L^2{\cal H}_M\approx1.1\times10^{33}\,{\rm Mx}^2\,{\rm cm}^{-1}$.
Several estimates of the gauge-invariant magnetic helicity of NOAA 11158
using time integration of photospheric magnetic helicity injection
\citep{VAMC12,LS12} and nonlinear force-free coronal field extrapolation
\citep{Jing12,Georg13} suggest magnetic helicities of the order of
$10^{43}\,{\rm Mx}^2$.
This value would be comparable to ours if the effective vertical
extent were $\approx100\,{\rm Mm}$.
We should remember, however, that there is no basis for such a vertical
extrapolation of our two-dimensional data.

Interestingly, the magnetic energy spectra $E_M^{(h)}(k)$ and $E_M^{(v)}(k)$
based respectively on the horizontal and vertical magnetic field components
agree remarkably well at wavenumbers below $k=3\,{\rm Mm}^{-1}$,
corresponding to length scales larger than 2\,Mm.
This suggests that our assumption of isotropy might be a reasonable one.
The mutual departure between $E_M^{(h)}(k)$ and $E_M^{(v)}(k)$
at larger wavenumbers could in principle be a physical effect,
although there is no good reason why the magnetic field 
should be mostly vertical only at small scales.
If it is indeed a physical effect, it should
then in future be possible to verify that this wavenumber,
where $E_M^{(h)}(k)$ and $E_M^{(v)}(k)$ depart from each other,
is independent of the instrument.
Alternatively, this departure might be connected with different accuracies
of horizontal and vertical magnetic field measurements \citep{Zhang2012}.
If that is the case, one should expect that with future measurements at better
resolution the two spectra depart from each other at larger wavenumbers.
In that case, our spectral analysis could be used to isolate potential
artefacts in the determination of horizontal and vertical magnetic fields.

In Figure~\ref{fig:heli_phelicity_Bran_zhq5_ene} we show
$2E_M(k)$ and $k|H_M(k)|$ for different days.
It turns out that on small scales the spectra are rather
similar in time, and that there are differences in the amplitude
mainly on large scales.
Also the sign of $H_M(k)$ remains positive for the different days.

We find that the mean spectral values of magnetic energy of
the active region at the solar surface is consistent with a $k^{-5/3}$
power law, which is expected based on the theory of \cite{GS95}
and consistent with spectra from earlier work on solar magnetic fields
\citep{Ab05,St12}, ruling out the $k^{-3/2}$ spectrum
suggested by \cite{Iro63} and \cite{Kra65}.

\begin{figure}[t!]
\begin{center}
\includegraphics[width=\columnwidth]{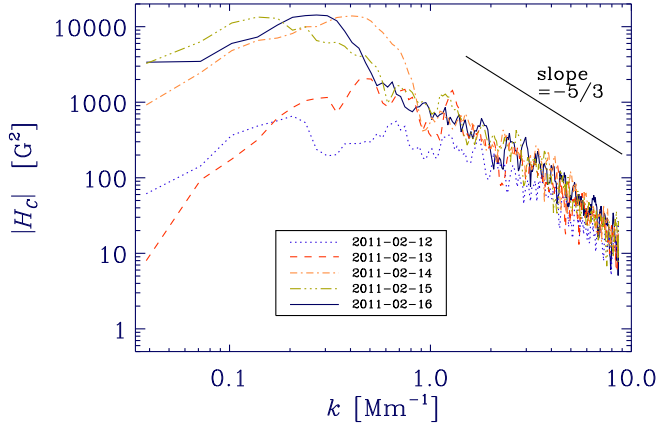}
\end{center}
\caption{Unsigned current helicity spectrum, $|H_C(k)|$.}
\label{fig:phc}
\end{figure}

Under isotropic conditions, the current helicity spectrum, $H_C(k,t)$,
is related to the magnetic helicity spectrum via \citep{Moffatt78}
\begin{equation}
H_C(k,t)\approx k^2H_M(k,t).\label{eq:helispec9}
\end{equation}
It is normalized such that $\int H_C(k)\,dk=\langle\bm{J}\cdot\bm{B}\rangle$.
In Figure~\ref{fig:phc} we show $|H_C(k)|$ obtained in this way.
For $k\ga1{\rm Mm}^{-1}$, the current helicity spectrum shows a $k^{-5/3}$
spectrum, which is consistent with numerical simulations of helically
forced hydromagnetic turbulence \citep{BS05b,Bra09}, and indicative of
a forward cascade of current helicity.
Similar spectra have also been obtained for the analogous case of kinetic helicity
\citep{AL77,BO97}.
These results imply that the relative helicity decreases
toward smaller scales; see the corresponding discussion on p.~286
of \cite{Moffatt78}.

\section{Conclusions}

We have applied a novel technique to estimate the magnetic helicity
spectrum using vector magnetogram data at the solar surface.
We have made use of the assumption that the spectral two-point
correlation tensor of the magnetic field can be approximated by its
isotropic representation.
This assumption is partially justified by the fact that the
energy spectra from horizontal and vertical magnetic fields
agree at wavenumbers below $2\,{\rm Mm}^{-1}$.
However, it will be important to assess the assumption of isotropy
in future work through comparison with simulations.
An example are the simulations of \cite{Losada}, who employed however
only a one-dimensional representation of the spectral two-point
correlation function.
Nevertheless, the present results look promising, because the sign of magnetic
helicity is the same over a broad range of wavenumbers and consistent
with that theoretically expected for the southern hemisphere.
This is consistent with the right-handed twist inferred from all
previous studies of NOAA 11158 using different methods.
Except for the smallest wavenumbers, magnetic and current helicities
have essentially the same sign.
Therefore, a sign change is only expected at smaller
wavenumbers corresponding to scales comparable to those of the Sun itself.

It would be useful to extend our analysis to a larger surface area of the
Sun to see whether there is evidence for a sign change toward small wavenumbers
and thus large scales reflecting the global magnetic field of the solar cycle.
Such a change of sign is expected from dynamo theory \citep{Bra01}
and is a consequence of the inverse cascade of magnetic helicity \citep{PFL}.
Figure~\ref{fig:phelicity} gives indications of an opposite sign for
$k\le0.1\,{\rm Mm}^{-1}$, which corresponds to scales that are still
much smaller than those of the Sun.
However, measurements of spectral power on scales comparable to those
of the observed magnetogram itself are not sufficiently reliable.

Our results suggest that the unsigned current helicity spectrum shows
a $k^{-5/3}$ power law.
This is in agreement with simulations of hydromagnetic turbulence
\citep{BS05b} and implies that the turbulence becomes progressively
less helical toward smaller scales.
Our results suggest that at a typical scale of $\ell_M\approx6\,$Mm,
the relative magnetic helicity reaches values around $0.04$.
This magnetic helicity must have its origin in the underlying
dynamo process, and can be traced back to the interaction between
rotation and stratification.
\cite{Losada} parameterized these two effects in terms of
a stratification parameter Gr and a Coriolis number Co and found
that the relative kinetic helicity is approximately 2\,Gr\,Co.
For the Sun, they estimate ${\rm Gr}=1/6.5$, so a relative helicity of
0.04 might correspond to ${\rm Co}\approx0.1$.
For the solar rotation rate, this corresponds to a correlation time
of about 6 hours, which translates to a depth of about 8\,Mm.
Again, more precise estimates should be obtained using realistic simulations.

In addition to measuring magnetic helicity over larger regions, it will
be important to apply our technique to many active regions covering both
hemispheres of the Sun and different times during the solar cycle.
This would allow us to verify the expected hemispheric dependence
of magnetic helicity.
Compared with previous determinations of the hemispheric
dependence of current helicity \citep{Zhang2012},
our technique might allow us to isolate instrumental artefacts
resulting from different resolutions of vector magnetograms
for horizontal and vertical magnetic fields.

\acknowledgments

We thank the referee for detailed and constructive comments
that have led to significant improvements of the manuscript.
This study is supported by grants from the
National Natural Science Foundation (NNSF) of China
under the project grants 10921303, 11221063 and 41174153 (HZ),
the NNSF of China and the Russian Foundation for Basic Research
under the collaborative China-Russian project 13-02-91158 (HZ+DDS),
the European Research Council under the
AstroDyn Research Project No.\ 227952, and
the Swedish Research Council under the project grants 2012-5797
and 621-2011-5076 (AB).

\newcommand{\yraa}[3]{ #1, {RAA,} {#2}, #3}
\newcommand{\yapj}[3]{ #1, {ApJ,} {#2}, #3}
\newcommand{\yjfm}[3]{ #1, {JFM,} {#2}, #3}
\newcommand{\ymn}[3]{ #1, {MNRAS,} {#2}, #3}
\newcommand{\yana}[3]{ #1, {A\&A,} {#2}, #3}
\newcommand{\yprl}[3]{ #1, {PhRvL,} {#2}, #3}
\newcommand{\yprd}[3]{ #1, {PhRvD,} {#2}, #3}
\newcommand{\ypre}[3]{ #1, {PhRvE,} {#2}, #3}
\newcommand{\yjswsc}[3]{ #1, {JSWJC,} {#2}, #3}
\newcommand{\ysov}[3]{ #1, {Sov.\ Astron.,} {#2}, #3}
\newcommand{\ypf}[3]{ #1, {Phys.\ Fluids,} {#2}, #3}
\newcommand{\ybook}[3]{ #1, {#2} (#3)}

\end{document}